\documentclass[universe,article,accept,moreauthors,pdftex]{Definitions/mdpi}

\usepackage[normalem]{ulem} 
\usepackage{amsfonts} 
\usepackage{wasysym} 
\usepackage{mathcomp} 
\usepackage{CJKutf8} 
\usepackage{pifont} 
\usepackage{bm} 
\usepackage{bbm} 
\graphicspath{{./Definitions/}} 

\makeatletter
\def\T@n@@nc@d@ngM@cr@M@d{}
\def\LY@n@@nc@d@ngM@cr@M@d{}
\makeatother

\let\orignewcommand\newcommand  
\let\newcommand\providecommand  
\usepackage{verse}
\let\newcommand\orignewcommand  

\newsavebox\foobox

\renewcommand{\dagger}{\mathchar"2279}
\renewcommand{\ddagger}{\mathchar"227A}

\newcommand{\mmathit}[1]{
  \ifthenelse{\equal{#1}{\ln}}{\mathit{ln}}{
    \ifthenelse{\equal{#1}{\max}}{\mathit{max}}{\mathit{#1}}
  }
}
\makeatother
\robustify{\footnote} 
  
\DeclareUnicodeCharacter{1E45}{\.{n}}
\DeclareUnicodeCharacter{1E41}{\.{m}}
\DeclareUnicodeCharacter{2003}{\quad}
\DeclareUnicodeCharacter{2009}{\thinspace}
\DeclareUnicodeCharacter{2002}{\enspace{}}
\DeclareUnicodeCharacter{2005}{\thinspace}
\DeclareUnicodeCharacter{0263}{\textipa{G}}
\DeclareUnicodeCharacter{A0}{~}
\DeclareUnicodeCharacter{2460}{\textcircled{\scriptsize{1}}}
\DeclareUnicodeCharacter{2461}{\textcircled{\scriptsize{2}}}
\DeclareUnicodeCharacter{2462}{\textcircled{\scriptsize{3}}}
\DeclareUnicodeCharacter{2463}{\textcircled{\scriptsize{4}}}
\DeclareUnicodeCharacter{2464}{\textcircled{\scriptsize{5}}}
\DeclareUnicodeCharacter{2465}{\textcircled{\scriptsize{6}}}
\DeclareUnicodeCharacter{2466}{\textcircled{\scriptsize{7}}}
\DeclareUnicodeCharacter{2467}{\textcircled{\scriptsize{8}}}
\DeclareUnicodeCharacter{2468}{\textcircled{\scriptsize{9}}}
\DeclareUnicodeCharacter{2070}{\textsuperscript{0}}
\DeclareUnicodeCharacter{2074}{\textsuperscript{4}}
\DeclareUnicodeCharacter{2075}{\textsuperscript{5}}
\DeclareUnicodeCharacter{2076}{\textsuperscript{6}}
\DeclareUnicodeCharacter{2077}{\textsuperscript{7}}
\DeclareUnicodeCharacter{2078}{\textsuperscript{8}}
\DeclareUnicodeCharacter{2079}{\textsuperscript{9}}
\DeclareUnicodeCharacter{02C2}{<}
\DeclareUnicodeCharacter{2033}{\relax\ifmmode '' \else $''$\fi}
\DeclareUnicodeCharacter{2034}{\relax\ifmmode ''' \else $'''$\fi}
\DeclareUnicodeCharacter{2026}{\relax\ifmmode … \else $\ldots$\fi}
\DeclareUnicodeCharacter{0229}{\c{e}}
\DeclareUnicodeCharacter{016F}{\r{u}}
\DeclareUnicodeCharacter{127}{\relax\ifmmode\rm\hbar\else $\rm\hbar$\fi}
\DeclareUnicodeCharacter{3AC}{\relax\ifmmode\acute{\alpha}\else $\acute{\alpha}$\fi}
\DeclareUnicodeCharacter{3AD}{\relax\ifmmode\acute{\varepsilon}\else $\acute{\varepsilon}$\fi}
\DeclareUnicodeCharacter{3AE}{\relax\ifmmode\acute{\eta}\else $\acute{\eta}$\fi}
\DeclareUnicodeCharacter{3AF}{\relax\ifmmode\acute{\iota}\else $\acute{\iota}$\fi}
\DeclareUnicodeCharacter{3CC}{\relax\ifmmode\acute{o}\else $\acute{o}$\fi}
\DeclareUnicodeCharacter{3CD}{\relax\ifmmode\acute{\upsilon}\else $\acute{\upsilon}$\fi}
\DeclareUnicodeCharacter{3CE}{\relax\ifmmode\acute{\omega}\else $\acute{\omega}$\fi}
\DeclareUnicodeCharacter{391}{A}
\DeclareUnicodeCharacter{392}{B}
\DeclareUnicodeCharacter{395}{E}
\DeclareUnicodeCharacter{396}{Z}
\DeclareUnicodeCharacter{397}{H}
\DeclareUnicodeCharacter{399}{I}
\DeclareUnicodeCharacter{39A}{K}
\DeclareUnicodeCharacter{39C}{M}
\DeclareUnicodeCharacter{39D}{N}
\DeclareUnicodeCharacter{39F}{O}
\DeclareUnicodeCharacter{3A1}{P}
\DeclareUnicodeCharacter{3A4}{T}
\DeclareUnicodeCharacter{3A7}{X}

\DeclareUnicodeCharacter{27E6}{\relax\ifmmode \llbracket \else $\llbracket$\fi}
\DeclareUnicodeCharacter{27E7}{\relax\ifmmode \rrbracket \else $\rrbracket$\fi}

\DeclareUnicodeCharacter{1D434}{\relax\ifmmode A \else $A$\fi}
\DeclareUnicodeCharacter{1D435}{\relax\ifmmode B \else $B$\fi}
\DeclareUnicodeCharacter{1D436}{\relax\ifmmode C \else $C$\fi}
\DeclareUnicodeCharacter{1D437}{\relax\ifmmode D \else $D$\fi}
\DeclareUnicodeCharacter{1D438}{\relax\ifmmode E \else $E$\fi}
\DeclareUnicodeCharacter{1D439}{\relax\ifmmode F \else $F$\fi}
\DeclareUnicodeCharacter{1D43A}{\relax\ifmmode G \else $G$\fi}
\DeclareUnicodeCharacter{1D43B}{\relax\ifmmode H \else $H$\fi}
\DeclareUnicodeCharacter{1D43C}{\relax\ifmmode I \else $I$\fi}
\DeclareUnicodeCharacter{1D43D}{\relax\ifmmode J \else $J$\fi}
\DeclareUnicodeCharacter{1D43E}{\relax\ifmmode K \else $K$\fi}
\DeclareUnicodeCharacter{1D43F}{\relax\ifmmode L \else $L$\fi}
\DeclareUnicodeCharacter{1D440}{\relax\ifmmode M \else $M$\fi}
\DeclareUnicodeCharacter{1D441}{\relax\ifmmode N \else $N$\fi}
\DeclareUnicodeCharacter{1D442}{\relax\ifmmode O \else $O$\fi}
\DeclareUnicodeCharacter{1D443}{\relax\ifmmode P \else $P$\fi}
\DeclareUnicodeCharacter{1D444}{\relax\ifmmode Q \else $Q$\fi}
\DeclareUnicodeCharacter{1D445}{\relax\ifmmode R \else $R$\fi}
\DeclareUnicodeCharacter{1D446}{\relax\ifmmode S \else $S$\fi}
\DeclareUnicodeCharacter{1D447}{\relax\ifmmode T \else $T$\fi}
\DeclareUnicodeCharacter{1D448}{\relax\ifmmode U \else $U$\fi}
\DeclareUnicodeCharacter{1D449}{\relax\ifmmode V \else $V$\fi}
\DeclareUnicodeCharacter{1D44A}{\relax\ifmmode W \else $W$\fi}
\DeclareUnicodeCharacter{1D44B}{\relax\ifmmode X \else $X$\fi}
\DeclareUnicodeCharacter{1D44C}{\relax\ifmmode Y \else $Y$\fi}
\DeclareUnicodeCharacter{1D44D}{\relax\ifmmode Z \else $Z$\fi}
\DeclareUnicodeCharacter{1D44E}{\relax\ifmmode a \else $a$\fi}
\DeclareUnicodeCharacter{1D44F}{\relax\ifmmode b \else $b$\fi}
\DeclareUnicodeCharacter{1D450}{\relax\ifmmode c \else $c$\fi}
\DeclareUnicodeCharacter{1D451}{\relax\ifmmode d \else $d$\fi}
\DeclareUnicodeCharacter{1D452}{\relax\ifmmode e \else $e$\fi}
\DeclareUnicodeCharacter{1D453}{\relax\ifmmode f \else $f$\fi}
\DeclareUnicodeCharacter{1D454}{\relax\ifmmode g \else $g$\fi}
\DeclareUnicodeCharacter{1D456}{\relax\ifmmode i \else $i$\fi}
\DeclareUnicodeCharacter{1D457}{\relax\ifmmode j \else $j$\fi}
\DeclareUnicodeCharacter{1D458}{\relax\ifmmode k \else $k$\fi}
\DeclareUnicodeCharacter{1D459}{\relax\ifmmode l \else $l$\fi}
\DeclareUnicodeCharacter{1D45A}{\relax\ifmmode m \else $m$\fi}
\DeclareUnicodeCharacter{1D45B}{\relax\ifmmode n \else $n$\fi}
\DeclareUnicodeCharacter{1D45C}{\relax\ifmmode o \else $o$\fi}
\DeclareUnicodeCharacter{1D45D}{\relax\ifmmode p \else $p$\fi}
\DeclareUnicodeCharacter{1D45E}{\relax\ifmmode q \else $q$\fi}
\DeclareUnicodeCharacter{1D45F}{\relax\ifmmode r \else $r$\fi}
\DeclareUnicodeCharacter{1D460}{\relax\ifmmode s \else $s$\fi}
\DeclareUnicodeCharacter{1D461}{\relax\ifmmode t \else $t$\fi}
\DeclareUnicodeCharacter{1D462}{\relax\ifmmode u \else $u$\fi}
\DeclareUnicodeCharacter{1D463}{\relax\ifmmode v \else $v$\fi}
\DeclareUnicodeCharacter{1D464}{\relax\ifmmode w \else $w$\fi}
\DeclareUnicodeCharacter{1D465}{\relax\ifmmode x \else $x$\fi}
\DeclareUnicodeCharacter{1D466}{\relax\ifmmode y \else $y$\fi}
\DeclareUnicodeCharacter{1D467}{\relax\ifmmode z \else $z$\fi}

\DeclareUnicodeCharacter{1E67}{\.{\v s}}
\DeclareUnicodeCharacter{1E11}{\relax\ifmmode \c{d} \else $\c{d}$\fi}
\DeclareUnicodeCharacter{1ECB}{\relax\ifmmode \d{i} \else $\d{i}$\fi}
\DeclareUnicodeCharacter{1D8D}{\relax\ifmmode \textlhookx \else $\textlhookx$\fi}
\DeclareUnicodeCharacter{104}{\relax\ifmmode \k{A} \else $\k{A}$\fi}
\DeclareUnicodeCharacter{211E}{\relax\ifmmode \textrecipe \else $\textrecipe$\fi}
\DeclareUnicodeCharacter{29D}{\relax\ifmmode \textctj \else $\textctj$\fi}

\DeclareUnicodeCharacter{1E2E}{\'{\"I}}
\DeclareUnicodeCharacter{23F}{\textrts}

\DeclareUnicodeCharacter{2C73}{\varw}

\DeclareUnicodeCharacter{2127}{\mho}

\DeclareUnicodeCharacter{28C}{\textturnv}
\DeclareUnicodeCharacter{252}{\textturnscripta}
\DeclareUnicodeCharacter{259}{\schwa}
\DeclareUnicodeCharacter{25B}{\m{e}}
\DeclareUnicodeCharacter{266}{\m{h}}
\DeclareUnicodeCharacter{127}{\B{h}}
\DeclareUnicodeCharacter{27E}{\textfishhookr}
\DeclareUnicodeCharacter{281}{\textinvscr}

\firstpage{1}
  \articlenumber{319}
\pubvolume{10}
\issuenum{8}
\pubyear{2024}
\copyrightyear{2024}
\externaleditor{Academic Editor: Yongquan~Xue}
\datereceived{12 June 2024}
\daterevised{25 July 2024}
\dateaccepted{5 August 2024}
\datepublished{8 August 2024}
\hreflink{https://doi.org/\linebreak10.3390/universe10080319}
\usepackage[OT1,OT2,T2A,T2B,T2C,T3,T5,T1]{fontenc}
\usepackage[russian,english]{babel}

\Title{Characteristics of Powerful Radio Galaxies}

\TitleCitation{Characteristics of Powerful Radio Galaxies}

\Author{Chandra B.~Singh~\textsuperscript{1}\textsuperscript{,}*, Michael~Williams~\textsuperscript{2}, David~Garofalo~\textsuperscript{2}\textsuperscript{,}*\href{https://orcid.org/0000-0001-5536-829X}{\orcidicon}, Luis~Rojas Castillo~\textsuperscript{2}, Landon~Taylor~\textsuperscript{2} and Eddie~Harmon~\textsuperscript{2}}

\AuthorNames{Chandra B. Singh, Michael Williams, David Garofalo, Luis Rojas Castillo, Landon Taylor and Eddie Harmon}

\AuthorCitation{Singh, C.B.; Williams, M.; Garofalo, D.; Rojas Castillo, L.; Taylor, L.; Harmon, E.}

\address{\textsuperscript{1} \quad South-Western Institute for Astronomy Research~(SWIFAR), Yunnan University, University Town, Chenggong, \mbox{Kunming 650500, China}

\textsuperscript{2} \quad Department of Physics, Kennesaw State University, Marietta, GA 30060, USA}

\corres{Correspondence: chandrasingh@ynu.edu.cn (C.B.S.); dgarofal@kennesaw.edu (D.G.)}

\abstract{Mature radio galaxies such as M87 belong to a specific subclass of active galaxies (AGN) whose evolution in time endows them with five distinguishing characteristics, including (1) low excitation emission, (2) low star formation rates, (3) high bulge stellar-velocity dispersion, (4) bright stellar nuclei, and (5) weak or nonexistent merger signatures. We show how to understand these seemingly disparate characteristics as originating from the time evolution of powerful radio quasars and describe a new model prediction that tilted accretion disks in AGN are expected to occur in bright quasars but not in other subclasses of AGN. The picture we present should be understood as the most compelling evidence for counter-rotation as a key element in feedback from accreting \mbox{black holes}.}


\makeatletter
\DeclareRobustCommand*\textsubscript[1]{%
  \@textsubscript{\selectfont#1}}
\def\@textsubscript#1{%
  {\m@th\ensuremath{_{\mbox{\fontsize\sf@size\z@#1}}}}}
\makeatother

\usepackage{sansmath}

\begin{document}
\section{Introduction \label{sect:sec1-universe-3078783}}

Understanding feeding and feedback from black holes has been a cornerstone of extragalactic astronomy since radio quasars were detected and interpreted in the 1960s~\mbox{\cite{B1-universe-3078783,B2-universe-3078783,B3-universe-3078783,B4-universe-3078783,B5-universe-3078783}}. Models for accreting black holes date back to the 1970s~\cite{B6-universe-3078783,B7-universe-3078783,B8-universe-3078783}, with those for AGN coming into their own by the 1990s~\cite{B9-universe-3078783,B10-universe-3078783,B11-universe-3078783}. Despite some success, the most widely accepted paradigm for astrophysical black holes has yet to explain a number of observations: AGN with jets are the minority; powerful jetted AGN live earlier, while bright, jetless AGN peak later; jetted AGN have more massive black holes; and merger signatures in jetted/non-jetted AGN do not unambiguously point to a trigger. FRI radio galaxies have the most massive black holes while FRII radio galaxies have less massive ones. FRII high-excitation radio galaxies (HERG) are at higher redshift, FRII low-excitation radio galaxies (LERG) are at intermediate redshift, and FRI LERG are at lower redshift. The key missing element that allows a simultaneous understanding of all the above is counter-rotation between accretion disks and black holes~\cite{B12-universe-3078783}. With this idea, we have not only addressed the above issues but have recently been able to explain the details of the M-$\upsigma$ relations between black hole mass and stellar velocity dispersion~\cite{B13-universe-3078783} and the lifetimes of radio quasar jets. This model is compatible with the jet/wind scenario discussed in~\cite{B14-universe-3078783}, as explained in~\cite{B15-universe-3078783}. Here, we connect the relation among star formation rate, stellar velocity dispersion, core stellar brightness, excitation level, and merger signatures. In \sect{sect:sec2dot1-universe-3078783}, we describe the data, and in \sect{sect:sec2dot2-universe-3078783}, we describe the theory that explains it. In \sect{sect:sec3-universe-3078783}, we conclude.

\section{Discussion \label{sect:sec2-universe-3078783}}

\subsection{Observations \label{sect:sec2dot1-universe-3078783}}

\begin{figure}[H]
      \includegraphics[scale=1]{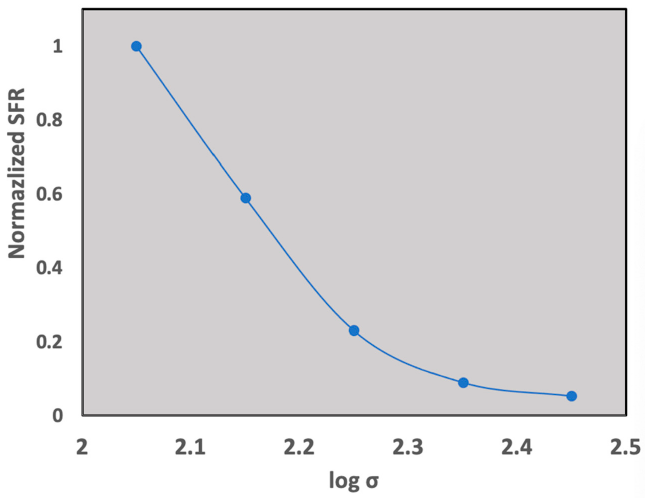}
\caption{z = 0.6--1.0 ratio of star-forming galaxy density to quiescent galaxy density constructed from the velocity dispersion functions in~\cite{B16-universe-3078783} versus stellar velocity dispersion as a proxy for the rate of star formation.}
\label{fig:universe-3078783-f001}
\end{figure}

In this section, we begin by deriving a proxy for the average---or overall---behavior of star formation rates as a function of galaxy-bulge stellar velocity dispersion at late times in the universe (i.e., past the peak of quasars at a redshift of 2). To accomplish this, we extract the densities of star-forming galaxies (\tabref{tabref:universe-3078783-t001}) and of quiescent galaxies (\tabref{tabref:universe-3078783-t002}) from the velocity dispersion functions (VDF) of~\cite{B16-universe-3078783}, which are given in terms of logarithms. If you take the antilogarithms of the density values in \cref{tabref:universe-3078783-t001,tabref:universe-3078783-t002}, i.e., 10\textsuperscript{log ${\upvarphi}$} where log ${\upvarphi}$ is given in~\cite{B16-universe-3078783}, you will reproduce the VDF of~\cite{B16-universe-3078783}. With these densities, we construct the ratio of star-forming galaxy density to quiescent galaxy density (\tabref{tabref:universe-3078783-t003}) and consider it as a proxy for the average rate of star formation between redshift of 0.6 and 1. We show the anti-correlation between \tabref{tabref:universe-3078783-t003} data and stellar velocity dispersion in \fig{fig:universe-3078783-f001}. This is strong evidence that galaxies with high stellar velocity dispersions tend to have low star formation rates~\cite{B17-universe-3078783,B18-universe-3078783}.    
    \begin{table}[H]
    \tablesize{\small}
    \caption{Density of star-forming galaxies from the velocity dispersion function of {\cite{B16-universe-3078783}}.}
    \label{tabref:universe-3078783-t001}

\setlength{\cellWidtha}{\textwidth/6-2\tabcolsep-0in}
\setlength{\cellWidthb}{\textwidth/6-2\tabcolsep-0in}
\setlength{\cellWidthc}{\textwidth/6-2\tabcolsep-0in}
\setlength{\cellWidthd}{\textwidth/6-2\tabcolsep-0in}
\setlength{\cellWidthe}{\textwidth/6-2\tabcolsep-0in}
\setlength{\cellWidthf}{\textwidth/6-2\tabcolsep-0in}
\scalebox{1}[1]{\begin{tabularx}{\textwidth}{>{\raggedright\arraybackslash}m{\cellWidtha}>{\raggedright\arraybackslash}m{\cellWidthb}>{\raggedright\arraybackslash}m{\cellWidthc}>{\raggedright\arraybackslash}m{\cellWidthd}>{\raggedright\arraybackslash}m{\cellWidthe}>{\raggedright\arraybackslash}m{\cellWidthf}}
\toprule

\textbf{log $\bm{\upsigma}$} & \textbf{z = 0.6--1} & \textbf{Z = 0.6--0.7} & \textbf{z = 0.7--0.8} & \textbf{z = 0.8--0.9} & \textbf{z = 0.9--1}\\
\cmidrule{1-6}

2.05 & 0.018 & 0.012 & 0.017 & 0.0195 & 0.028\\
\cmidrule{1-6}
2.15 & 0.011 & 0.007 & 0.011 & 0.0123 & 0.016\\
\cmidrule{1-6}
2.25 & 0.004 & 0.0026 & 0.002 & 0.007 & 0.007\\
\cmidrule{1-6}
2.35 & 0.0008 & 0.00047 & 0.0006 & 0.0007 & 0.001\\
\cmidrule{1-6}
2.45 & 0.0001 & 0.0001 & 8.9125E-05 & 0.00013 & 0.0002\\

\bottomrule
\end{tabularx}}

    \end{table}
    \vspace{-9pt}
    
    \begin{table}[H]
    \tablesize{\small}
    \caption{Density of quiescent galaxies from the velocity dispersion function of {\cite{B16-universe-3078783}}.}
    \label{tabref:universe-3078783-t002}

\setlength{\cellWidtha}{\textwidth/6-2\tabcolsep-0in}
\setlength{\cellWidthb}{\textwidth/6-2\tabcolsep-0in}
\setlength{\cellWidthc}{\textwidth/6-2\tabcolsep-0in}
\setlength{\cellWidthd}{\textwidth/6-2\tabcolsep-0in}
\setlength{\cellWidthe}{\textwidth/6-2\tabcolsep-0in}
\setlength{\cellWidthf}{\textwidth/6-2\tabcolsep-0in}
\scalebox{1}[1]{\begin{tabularx}{\textwidth}{>{\raggedright\arraybackslash}m{\cellWidtha}>{\raggedright\arraybackslash}m{\cellWidthb}>{\raggedright\arraybackslash}m{\cellWidthc}>{\raggedright\arraybackslash}m{\cellWidthd}>{\raggedright\arraybackslash}m{\cellWidthe}>{\raggedright\arraybackslash}m{\cellWidthf}}
\toprule

\textbf{log $\bm{\upsigma}$} & \textbf{z = 0.6--1} & \textbf{z = 0.6--0.7} & \textbf{z = 0.7--0.8} & \textbf{z = 0.8--0.9} & \textbf{z = 0.9--1}\\
\cmidrule{1-6}

2.04 & 0.005 & 0.005 & 0.009 & 0.005 &  \\
\cmidrule{1-6}
2.15 & 0.006 & 0.0044 & 0.006 & 0.007 & 0.007\\
\cmidrule{1-6}
2.25 & 0.0054 & 0.0049 & 0.005 & 0.007 & 0.003\\
\cmidrule{1-6}
2.35 & 0.003 & 0.0025 & 0.002 & 0.003 & 0.003\\
\cmidrule{1-6}
2.45 & 0.0008 & 0.00071 & 0.0005 & 0.0007 & 0.001\\

\bottomrule
\end{tabularx}}

    \end{table}
    \vspace{-9pt}
    
    \begin{table}[H]
    \tablesize{\small}
    \caption{Ratio of data in \cref{tabref:universe-3078783-t001,tabref:universe-3078783-t002}.}
    \label{tabref:universe-3078783-t003}

\setlength{\cellWidtha}{\textwidth/5-2\tabcolsep-0in}
\setlength{\cellWidthb}{\textwidth/5-2\tabcolsep-0in}
\setlength{\cellWidthc}{\textwidth/5-2\tabcolsep-0in}
\setlength{\cellWidthd}{\textwidth/5-2\tabcolsep-0in}
\setlength{\cellWidthe}{\textwidth/5-2\tabcolsep-0in}
\scalebox{1}[1]{\begin{tabularx}{\textwidth}{>{\raggedright\arraybackslash}m{\cellWidtha}>{\raggedright\arraybackslash}m{\cellWidthb}>{\raggedright\arraybackslash}m{\cellWidthc}>{\raggedright\arraybackslash}m{\cellWidthd}>{\raggedright\arraybackslash}m{\cellWidthe}}
\toprule

\textbf{z = 0.6--1} & \textbf{z = 0.6--0.7} & \textbf{z = 0.7--0.8} & \textbf{z = 0.8--0.9} & \textbf{z = 0.9--1}\\
\cmidrule{1-5}

3.39 & 2.34 & 1.8 & 4.27 &  \\
\cmidrule{1-5}
2.0 & 1.59 & 1.9 & 1.82 & 2.1\\
\cmidrule{1-5}
0.78 & 0.52 & 0.49 & 1 & 2.04\\
\cmidrule{1-5}
0.302 & 0.19 & 0.34 & 0.28 & 0.42\\
\cmidrule{1-5}
0.18 & 0.16 & 0.19 & 0.18 & 0.18\\

\bottomrule
\end{tabularx}}

    \end{table}

In addition to the inverse relation between star formation rate and stellar velocity dispersion, the evidence associates low star formation rates with the jetted AGN subgroup and higher star formation rates with the non-jetted AGN subgroup.~\cite{B19-universe-3078783} evaluated the median star formation rate for MaNGA, with both jetted and non-jetted AGN, finding the non-jetted AGN to have a median star formation rate of 0.631 solar masses per year while the jetted AGN have a median star formation rate of 0.025. This is a factor of 25 difference in the star formation rate between the two AGN subclasses for \mbox{0.01 \textless{} z \textless{} 0.15}. The star formation rates for the radio quiet quasars of~\cite{B20-universe-3078783} are between 20 and 300 solar masses/year.~\cite{B21-universe-3078783} found their sample of radio quiet quasars to have an average star formation rate of 30 solar masses/year for 160 radio-quiet quasars.
    
In addition to star formation and velocity dispersion, we show evidence for a larger stellar component in the nucleus of powerful, low-excitation FRI radio galaxies (i.e., exhibiting weak emission lines) as compared to radio-quiet quasars. We look at the data from~\cite{B22-universe-3078783} and pick out the elliptical galaxies in their sample with the largest Sersic indices and with jets and report them in \tabref{tabref:universe-3078783-t004}. The average Sersic index is n = 9.7. The average Sersic index for the disk galaxies of~\cite{B22-universe-3078783} is 3.3. n, which appears in the intensity profile as a function of distance r as (see~\cite{B23-universe-3078783} for details)
        \begin{equation}
\nonumber\nonumber
\text{I(r) = I\textsubscript{e}exp \{$-$b\textsubscript{n}[(r/r\textsubscript{e})\textsuperscript{1/n} $-$ 1]\}.}
\end{equation}
    
    \begin{table}[H]
    \tablesize{\small}
    \caption{1. Object, 2. black hole mass in solar masses, 3. jet type, 4. accretion rate in terms of the Eddington limit, 5. star formation in terms of solar masses per year, 6. stellar velocity dispersion in km/s, and 7. sersic index.}
    \label{tabref:universe-3078783-t004}

\setlength{\cellWidtha}{\textwidth/7-2\tabcolsep-0in}
\setlength{\cellWidthb}{\textwidth/7-2\tabcolsep+0.4in}
\setlength{\cellWidthc}{\textwidth/7-2\tabcolsep-0in}
\setlength{\cellWidthd}{\textwidth/7-2\tabcolsep+0.4in}
\setlength{\cellWidthe}{\textwidth/7-2\tabcolsep-0.2in}
\setlength{\cellWidthf}{\textwidth/7-2\tabcolsep-0.3in}
\setlength{\cellWidthg}{\textwidth/7-2\tabcolsep-0.3in}
\scalebox{1}[1]{\begin{tabularx}{\textwidth}{>{\centering\arraybackslash}m{\cellWidtha}>{\centering\arraybackslash}m{\cellWidthb}>{\centering\arraybackslash}m{\cellWidthc}>{\centering\arraybackslash}m{\cellWidthd}>{\centering\arraybackslash}m{\cellWidthe}>{\centering\arraybackslash}m{\cellWidthf}>{\centering\arraybackslash}m{\cellWidthg}}
\toprule

\textbf{Object} & \textbf{Black Hole Mass} & \textbf{Jet Type} & \textbf{Accretion Rate} & \textbf{SFR} & \textbf{$\bm{\upsigma}$} & \textbf{n}\\
\cmidrule{1-7}

NGC 1399 & 5 $\times$ 10\textsuperscript{8}  & FRI & 9.1 $\times$ 10\textsuperscript{$-$3} & low &   & 16.8\\
\cmidrule{1-7}
NGC 4261 & 1.67 billion & FRI & 6 $\times$ 10\textsuperscript{$-$2} & low & 297 & 7.3\\
\cmidrule{1-7}
NGC 4374 & 0.93 billion & FRI & 2.4 $\times$ 10\textsuperscript{$-$3} & low & 278 & 5.6\\
\cmidrule{1-7}
NGC 4486 & 6.5 billion & FRI & 0.13 & low & 323 & 6.86\\

\bottomrule
\end{tabularx}}

    \end{table}

NGC 1399, the dominant galaxy in the Fornax Cluster, has a black hole just under a billion solar masses, an FRI jet, an estimated Bondi accretion rate of 9.1 $\times$ 10\textsuperscript{$-$3} solar masses/year~\cite{B24-universe-3078783}, and a low rate of star formation~\cite{B25-universe-3078783}. NGC 4261 is an elliptical galaxy in the Virgo Cluster with a black hole about 1.67 billion solar masses~\cite{B23-universe-3078783}, a visible two-sided FRI jet~\cite{B26-universe-3078783,B27-universe-3078783}, a low star formation rate~\cite{B28-universe-3078783}, a stellar velocity dispersion of 297 km/s~\cite{B24-universe-3078783}, and a low excitation nucleus~\cite{B29-universe-3078783}, with an estimated Bondi accretion rate of 6 $\times$ 10\textsuperscript{$-$2} solar masses per year~\cite{B24-universe-3078783}. NGC 4374 is an elliptical galaxy in the Virgo Cluster with a black hole of 0.93 billion solar masses~\cite{B24-universe-3078783}, a low rate of star formation~\cite{B28-universe-3078783,B30-universe-3078783}, an estimated Bondi accretion rate of 2.4 $\times$ 10\textsuperscript{$-$3} solar masses per year~\cite{B24-universe-3078783}, and a stellar velocity dispersion of 278 km/s~\cite{B24-universe-3078783}. NGC 4486 has a low star formation rate~\cite{B26-universe-3078783}, an estimated Bondi accretion rate of 0.13 solar masses per year~\cite{B24-universe-3078783}, and a stellar velocity dispersion of 323 km/s~\cite{B24-universe-3078783}. We do not discuss NGC 6251 with a Sersic index of 11.8 because it appears to be a giant radio galaxy, which is modeled in our paradigm as a system that experiences a second bout of counter-rotation. It also preserves its nuclear stellar component but requires further discussion beyond this work. Radio-quiet quasars, by comparison, appear to have higher star formation rates, lower velocity dispersions, higher excitation signatures, and dimmer nuclear stellar cores. The star formation rate of the radio-quiet quasar SBS 1520 + 530, for example, is measured to be 190 solar masses/year~\cite{B20-universe-3078783}, while its Sersic index is only 2.7~\cite{B31-universe-3078783}. Finally, the merger signatures in FRI radio galaxies are fewer than in other AGN~\cite{B32-universe-3078783}, and even more intriguing is the finding that the signatures of mergers decrease with increasing galactic stellar velocity dispersion~\cite{B33-universe-3078783}. We place these objects and their properties in \tabref{tabref:universe-3078783-t004}. In the next section, we describe the theoretical context that allows for an understanding of how these five characteristics come about in this particular subclass of AGN.

\subsection{Theory \label{sect:sec2dot2-universe-3078783}}

In this section, we use a simple model schematic to show how the formation and evolution of radio quasars strings together the five characteristics described in the previous section, namely,
        \begin{enumerate}[label=$\bullet$]
\item bright nuclear stellar cores
\item low excitation accretion
\item high stellar velocity dispersion
\item low star formation rates
\item weak merger signatures
\end{enumerate}

\begin{enumerate}
\item[\emph{a.} ] bright nuclear stellar cores
\end{enumerate}

\begin{figure}[H]
          \begin{adjustwidth}{-\extralength}{0cm}
      \centering
      \includegraphics[scale=0.9]{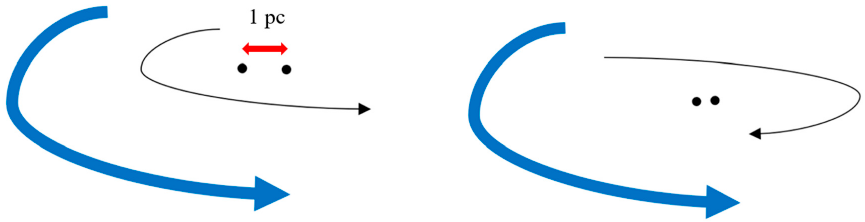}
          \end{adjustwidth}
\caption{(\textbf{\boldmath{Left}})---Merging black holes stuck at a 1 pc separation~\cite{B34-universe-3078783}. Stars are needed to extract the angular momentum of the black hole binary because its angular momentum is in the same direction as that of the incoming gas. Gas rotation indicated by thick blue arrow. Binary rotation indicated by thin black arrow. (\textbf{\boldmath{Right}})---Stars not needed to extract the angular momentum of the black hole binary. The counter-rotating disk is efficient in coalescing the black holes~\cite{B35-universe-3078783}, which are shown within 1 pc and heading toward coalescence.}
\label{fig:universe-3078783-f002}
\end{figure}

        Our theoretical picture is anchored to the idea of counter-rotation between a spinning black hole and an accretion disk, a configuration that results from a merger of two galaxies. Each galaxy provides a supermassive black hole that eventually coalesces. In order for this to occur, the black hole binary must rid itself of its binary angular momentum such that the black holes are well within a pc of each other, where gravitational waves can complete the merger. Because a corotating disk struggles to bring the binary within 1 pc~\cite{B34-universe-3078783}, nearby stars are needed to further eliminate the binary angular momentum, and this leads to a depleted stellar core. If the disk is counter-rotating with respect to the binary black hole angular momentum, on the other hand, the accretion disk is effective in extracting the binary angular momentum~\cite{B35-universe-3078783}, and the black holes can merge without depleting their stellar cores. This is illustrated in \fig{fig:universe-3078783-f002}. As we will show, the active galaxies triggered by counter-rotating accretion will tend to become the red-and-dead FRI radio galaxies which are therefore prescribed by theory to have bright nuclear cores.
		  \begin{enumerate}
\item[\emph{b.} ] low excitation accretion
\end{enumerate}

\begin{figure}[H]
      \includegraphics[scale=0.88]{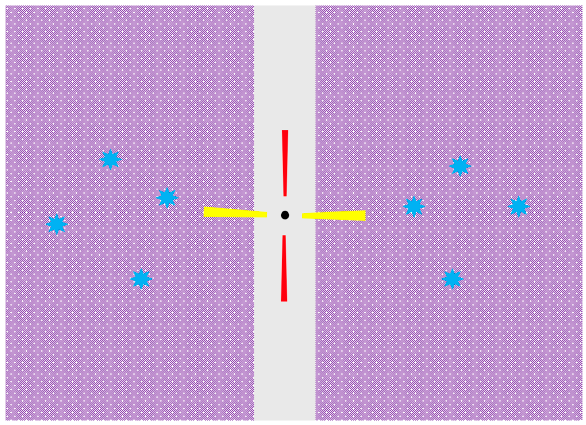}
\caption{A counter-rotating accretion disk (yellow) around a spinning black hole produces a jet (red) that pushes cold gas away from a funnel-like central region, leading to an enhancement in the star formation (blue).}
\label{fig:universe-3078783-f003}
\end{figure}

\begin{figure}[H]
      \includegraphics[scale=0.88]{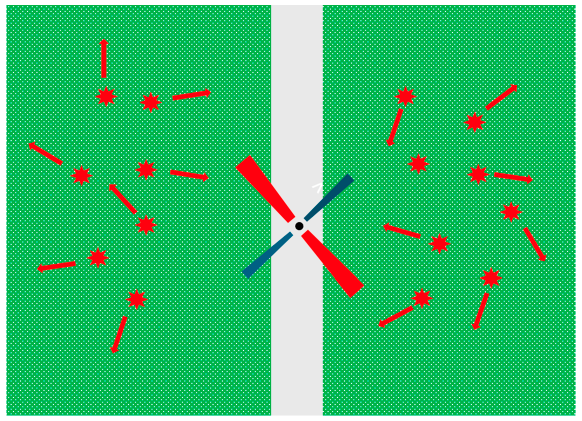}
\caption{In the transition through zero black-hole spin, a new plane of accretion is formed, and once the spin is large enough, a new jet emerges, which impinges on a higher-density region and is entrained, heating the gas, shutting down star formation, and enhancing stellar velocity dispersion (red arrows). The halo becomes suffused in hot gas (green), the star formation drops, and the stars become old (red). Note the transition from a radiatively efficient disk (bright yellow) to an ADAF (dull blue). This is a low-excitation radio galaxy. The gas is heated by the jet and, as a result, is suffused in a hot X-ray halo that cannot form stars.}
\label{fig:universe-3078783-f004}
\end{figure}

The counter-rotating accretion configuration is associated with an FRII jet~\cite{B12-universe-3078783}, whose effect is to create a funnel-like region through the ISM by pushing the cold gas away and enhancing the density. The higher-density region is shown in purple, while the less-dense funnel region is in grey in \fig{fig:universe-3078783-f003}. The energy deposited in the hotspots circles back to affect the accretion process by feeding heated gas and altering the state of accretion to an advection-dominated accretion flow (ADAF). Such a transition requires no less than \mbox{5 million years} and can occur during counter-rotation, but it dominates during the longest-lasting phase of such a radio galaxy, which follows the transition through zero black-hole spin~\cite{B12-universe-3078783}. This is captured by the change in color for the accretion disk from yellow in \fig{fig:universe-3078783-f003} to dirty blue in \fig{fig:universe-3078783-f004}.

\vspace{-12pt}

\vspace{-9pt}
    
		\begin{enumerate}
\item[\emph{c.} ] High stellar velocity dispersion
\end{enumerate}

\begin{figure}[H]
      \includegraphics[scale=1]{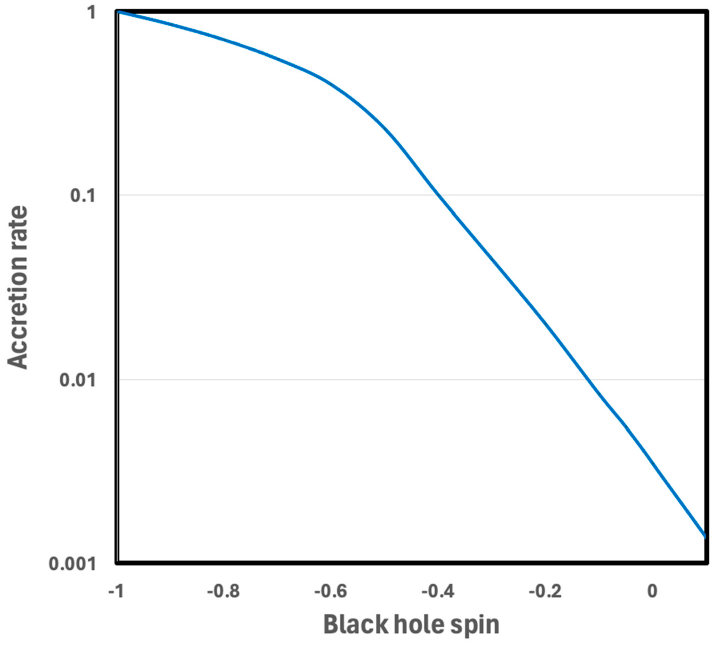}
\caption{Accretion rate onto the black hole in units of the Eddington accretion rate as a function of black hole spin. Evolution in time is toward zero black hole spin. This is the characteristic behavior for an AGN triggered into a radio quasar, whose disk evolves rapidly into an ADAF from a thin disk. The thin disk/ADAF boundary is located at 0.01 the Eddington accretion rate.}
\label{fig:universe-3078783-f005}
\end{figure}

\begin{figure}[H]
      \includegraphics[scale=1]{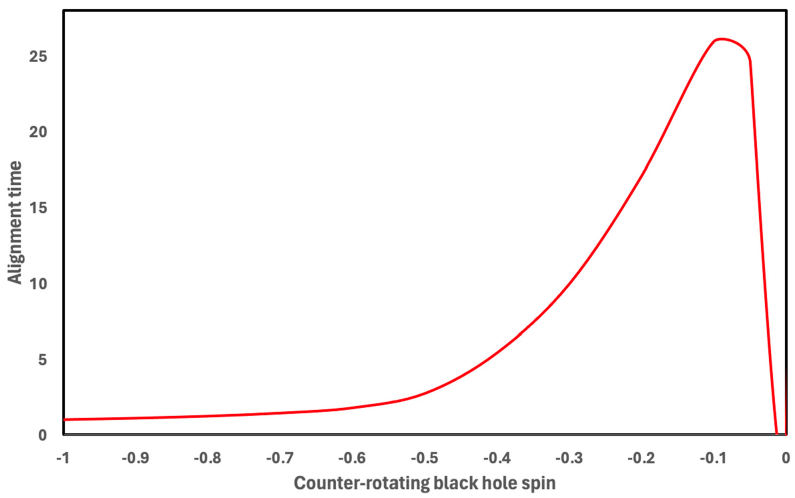}
\caption{Disk alignment timescale for a radio galaxy whose accretion rate reaches the thin-disk/ADAF boundary at a counter-rotating black hole spin of 0.1. The time is scaled to the initial timescale associated with an accretion rate that is at the Eddington limit.}
\label{fig:universe-3078783-f006}
\end{figure}

While the black hole spin has a non-zero value, the Bardeen--Petterson effect~\cite{B36-universe-3078783} applies, but it disappears when the black hole spin is zero and a new plane of accretion is formed that is determined by the angular momentum of the incoming gas~\cite{B37-universe-3078783} and which is responsible for the high stellar velocity dispersion as we now describe. The Bardeen--Petterson effect results from Lens--Thirring precession~\cite{B38-universe-3078783}, a general relativistic effect that ultimately forces the disk’s angular momentum to be aligned or anti-aligned with the angular momentum of the black hole. While the Bardeen--Petterson effect is believed to operate continuously in X-ray binaries, where the donor star feeds the compact object with angular momentum that is misaligned with that of the compact object, such an effect operates in a restrictive manner in the subset of AGN that lead to tilted jets, as we now show. The timescale for alignment is given by~\cite{B39-universe-3078783} as
        \begin{equation}
\nonumber\label{eq:FD1-universe-3078783}
T_{align} = \frac{3aM\left( {2R_{GR}/R_{BP}} \right)^{1/2}}{dM/dt}
\end{equation}
        where \emph{a} is the dimensionless black hole spin, M is the black hole mass, R\textsubscript{GR} is the gravitational radius, and R\textsubscript{BP} is the Bardeen--Petterson radius given as a function of black hole spin as
        \begin{equation}
\nonumber\nonumber
\text{\emph{R}\textsubscript{\emph{BP}} = \emph{Aa}\textsuperscript{2/3}\emph{R}\textsubscript{\emph{GR}}}
\end{equation}
        where A is constant, and dM/dt is the accretion rate onto the black hole. While this expression was derived for corotation, it has been shown that the timescale for counter-rotation is the same~\cite{B40-universe-3078783,B41-universe-3078783,B42-universe-3078783}. This expression tends to zero as the black hole spin approaches zero. According to our model prescription (\cite{B12-universe-3078783} and \fig{fig:universe-3078783-f004}), however, the accretion rate drops as the transition through zero spin occurs, which has the effect of increasing the alignment timescale. We model the timescale for an accreting black hole that approaches the boundary between thin disk and ADAF prior to reaching zero spin (\fig{fig:universe-3078783-f005}). As the black hole approaches zero spin in the counter-rotating regime, it initially experiences a jump in the alignment timescale due to the decrease in accretion rate (\fig{fig:universe-3078783-f006}), which is a factor of about 25 times larger than the alignment timescale of the post-merger black hole that accretes at the Eddington rate. A counter-rotating black hole spins down to zero spin from an initially maximally spinning black hole in about 8 $\times$ 10\textsuperscript{6} years if it accretes at the Eddington limit. Because the alignment timescale is longer than this~\cite{B39-universe-3078783}, counter-rotating black holes tend to have inner disks that are anti-aligned with the black hole angular momentum and outer disks beyond the Bardeen--Petterson radius that are tilted. This will last until the black hole is close to zero black hole spin. Once zero spin is reached, the Bardeen--Petterson effect disappears, the accretion disk forms in the plane determined by the angular momentum of the incoming gas from the greater galaxy, and the alignment timescale becomes irrelevant beyond that time. We therefore only show the alignment timescale as a function of spin during the counter-rotating regime. This bump in the alignment time is larger for the most powerful radio quasars, whose feedback effect rapidly transitions their thin disks into ADAFs. We therefore expect this effect to dominate in denser environments. Our model thus prescribes that FRII radio quasars/radio galaxies possess Bardeen--Petterson transition radii and that FRI radio galaxies do not. In other words, only in post-mergers will the accreting black holes have this feature. While this applies to radio-quiet quasars as well, we are here concerned with the jetted AGN subclass. Because radio quasars are modeled as counter-rotating black holes, they are relatively shorter-lived, at least compared to FRI radio galaxies. The observational consequences of this prediction remain unexplored.    
\vspace{-6pt}
    
When the black hole spins up again beyond the zero black hole spin in corotation and a jet is again formed, it will be tilted with respect to the earlier FRII jet phase, as shown in \fig{fig:universe-3078783-f004}. This has a direct impact on the stars in the bulge, which are pushed around and acquire an enhancement in their velocity dispersion (shown by red arrows). Note that the jet is drawn thicker as compared to the jet shown in \fig{fig:universe-3078783-f003}. This is due to the fact that the jet in \fig{fig:universe-3078783-f004} enters a denser region as compared to the FRII jet of \fig{fig:universe-3078783-f003}. Here we see an explanation for the different jet morphology. The FRI jet morphology is due to both an environmental effect (greater density) and an engine-based effect (its tilt).
		\begin{enumerate}
\item[\emph{d.} ] low star formation rates
\end{enumerate}

The absence of the Bardeen--Petterson effect and the tilted jet that develops in corotation will make it difficult to drill through the denser medium (\fig{fig:universe-3078783-f004}), and this leads to the FRI morphology, which deposits its energy in the ISM, heating the gas and impeding star formation. This effect on star formation is referred to as the Roy Conjecture~\cite{B43-universe-3078783}. As the star formation rate drops, the galaxy experiences increasingly fewer new stars and an aging of the existing stars, which gives the galaxy the character of an old red-and-dead elliptical galaxy. The old stars are drawn in red in \fig{fig:universe-3078783-f004}, and the hot halo is shown in green.
		\begin{enumerate}
\item[\emph{e.} ] weak merger signatures
\end{enumerate}

A powerful FRII jet that rapidly transitions its accretion disk to an ADAF in a few million years will accrete at 10\textsuperscript{$-$2} the Eddington limit, which means that it will spin its black hole down to zero spin in about 5 $\times$ 10\textsuperscript{8} years. In order to become an FRI radio galaxy, the black hole spin needs to reach a spin value of about 0.2, which, at the Eddington limit, takes about 20 million years. The accretion rate is at 10\textsuperscript{$-$2} the Eddington limit, however, and continues to decrease over this time period, so the FRI jet comes into play no sooner than about 2 $\times$ 10\textsuperscript{9} years after zero spin and lives as long as there is accretion fuel. Because of the low accretion rates at this late phase, FRI radio galaxies live billions of years in such states. Because merger signatures last about 100 million years, it is clear that for mature FRI radio galaxies like M87, no merger signatures are expected to survive. In general, the longer the FRI radio galaxy lives, the weaker the merger signatures become. Because theory prescribes jet affecting stellar velocity dispersion, the longer the FRI jet exists, the larger the predicted stellar velocity dispersions. Therefore, the model makes the interesting prediction of an inverse relation between stellar velocity dispersion and merger signatures. If galactic velocity dispersion increases with time, an anti-correlation is generated between merger signatures and galactic stellar velocity dispersion as is found observationally~\cite{B33-universe-3078783}.

\section{Conclusions \label{sect:sec3-universe-3078783}}

The fact that star formation is either enhanced~\cite{B44-universe-3078783,B45-universe-3078783} or suppressed~\cite{B46-universe-3078783,B47-universe-3078783,B48-universe-3078783,B49-universe-3078783,B50-universe-3078783} due to AGN feedback has been met with mixed evidence. Our paradigm for black hole feedback posits the enhancement of star formation for FRII radio galaxies associated with counter-rotating black holes. Jets likely tilt with respect to the radio quasar phase in the transition through zero spin to produce corotation, and this allows the jet energy to more directly couple to the interstellar medium, heating it and suppressing star formation. This tilt also affects stellar dispersions in the galactic bulge. Hence, in our paradigm, star formation suppression and the enhancement of stellar velocity dispersion go together. It is therefore of primary relevance to explore their connection over cosmic time.~\cite{B16-universe-3078783} have provided the data for this up to a redshift of 1. Galaxies with higher stellar dispersion values are strongly associated with lower star formation rates. Because the radio galaxies with the highest stellar dispersion values and lowest star formation rates are prescribed to be the end state of what originate as powerful radio quasars, a prediction is identified from the model that at high redshift, both suppressed star formation rates and stellar velocity dispersions should exhibit less-extreme values. While star formation rates should on average be higher than at lower redshift, the opposite would be true for stellar dispersions. Because this anti-correlation is instantiated as a result of original counter-rotation, black holes that are not triggered into counter-rotation will not experience it. Because counter-rotation is more efficient in merging black hole binaries, the AGN triggered in counter-rotation should experience non-depleted stellar cores. We have provided evidence of such correlations. In addition, both the low excitation character of a subset of radio galaxies and the weak merger signatures are a natural consequence of the late time evolution of these powerful AGNs. Finally, we have identified a prediction concerning the existence of tilted disks in a subset of AGN, whose observational features remain to be explored. Along with recent FRII jet lifetime estimates~\cite{B51-universe-3078783}, the evidence presented here constitutes strong support for counter-rotating black holes as a key element in our understanding of black hole feedback.

\vspace{6pt}
\authorcontributions{Conceptualization, C.B.S. and D.G.; methodology, D.G., L.R.C., L.T. and E.H.; software, M.W., L.R.C., L.T. and E.H.; formal analysis, M.W. and D.G.; writing---original draft, C.B.S. and D.G.; writing---review and editing, C.B.S. and D.G.; supervision, D.G. All authors have read and agreed to the published version of the manuscript.}
\funding{This research was funded by the National Natural Science Foundation of China, grant number 12073021.}
\dataavailability{No new data were created or analyzed in this study. Date sharing is not applicable to this article.}
\conflictsofinterest{The authors declare no conflicts of interest.}
\begin{adjustwidth}{-\extralength}{0cm}

\reftitle{References}

\end{adjustwidth}
\begin{adjustwidth}{-\extralength}{0cm}
\PublishersNote{}
\end{adjustwidth}

\end{document}